\def\year{2021}\relax
\documentclass[letterpaper]{article} 
\usepackage{aaai21}  
\usepackage{times}  
\usepackage{helvet} 
\usepackage{courier}  
\usepackage[hyphens]{url}  
\usepackage{graphicx} 
\usepackage{natbib}  
\usepackage{caption} 
\usepackage{algorithm}
\usepackage{algpseudocode}

\title{Object Recognition for Economic Development from Daytime Satellite Imagery}

\author {
        Klaus Ackermann,\textsuperscript{\rm 1,2*}
        Alexey Chernikov,\textsuperscript{\rm 1,3*}
        Nandini Anantharama,\textsuperscript{\rm 1,3} \\
        Miethy Zaman,\textsuperscript{\rm 1}
        Paul A~Raschky\textsuperscript{\rm 1,4}\\
}

\affiliations {
    \textsuperscript{\rm 1} SoDa Labs, Monash Business School, Monash University \\
    \textsuperscript{\rm 2} Department of Econometrics and Business Statistics, Monash University \\
    \textsuperscript{\rm 3} Faculty of Information Technology, Monash University \\
    \textsuperscript{\rm 4} Department of Economics, Monash University

    \{klaus.ackermann, alexey.chernikov, nandini.anantharama, miethy.zaman, paul.raschky\}@monash.edu
}
\begin{document}

\maketitle

\begin{abstract}
Reliable data about the stock of physical capital and infrastructure in developing countries is typically very scarce. This is particular a problem for data at the subnational level where existing data is often outdated, not consistently measured or coverage is incomplete. Traditional data collection methods are time and labor-intensive costly which often prohibits developing countries from collecting this type of data. This paper proposes a novel method to extract infrastructure features from high-resolution satellite images. We collected high-resolution satellite images for 5 million 1km $\times$ 1km grid cells covering 21 African countries. We contribute to the growing body of literature in this area by training our machine learning algorithm on ground-truth data. We show that our approach strongly improves the predictive accuracy. Our methodology can build the foundation to then predict subnational indicators of economic development for areas where this data is either missing or unreliable.
\end{abstract}

\section{Introduction}

The efficient allocation of limited governmental funds from local governments as well as international aid organizations crucially depends on reliable information about the level of socioeconomic indicators. These indicators (e.g. income, education, physical infrastructures, social class etc.) are critical inputs for addressing the socioeconomic issues for researchers and policy-makers alike. Although data availability and quality for the developing countries has been improving in recent years, consistently measured and reliable data is still relatively scarce. Numerous studies have documented specifically the problems of aggregate economic accounts, in particular to Africa, where the data suffers from various conceptual problems, measurement biases, and other errors  \citep[e.g.][]{Chen2011,Johnson2013,Jerven15}. 



Researchers have probed into alternative options in the absence of reliable official statistics. Among this newer generation of alternative economic data research, a burgeoning literature has emerged that uses satellite imagery of nighttime luminosity as a proxy for economic activity. Work by \citet{Sutton02}, \citet{Elvidge2009}, \citet{Chen2011}, \citet{Henderson2012}, \citet{Sutton07} and \citet{Hodler2014} documents a strong relationship between nighttime luminosity and gross domestic product (GDP) at the national and subnational levels. This allows researchers to generate information for any levels of regional analysis and also the likelihood of strategic, human manipulation is limited with satellite generated data. However, luminosity data as a proxy for economic activity is not free from concerns. Satellite sensors have a lower detection bound and nighttime light emissions below this bound are not captured by the satellites' readings. This leads to bottom-coding problem and this is particularly an issue in low-output and low-density regions \cite{Chen2011}, which are very often regions and countries (e.g. Africa) where official macroeconomic data is missing or unreliable as well.

Over the past few decades, some parts of the African continent have witnessed large increases in economic development. Nevertheless, the majority of regions within African nations still lacks behind. The continent faces further challenges due to localized conflicts \cite{berman2017}, rapid urbanization \cite{Aylin20} as well as the impacts of the COVID-19 pandemic \cite{Tregor20}, among others. A key pre-requisite in formulating adequate strategies to address these challenges, is reliable socioeconomic data at a spatially, granular level. As of now, even data about basic infrastructure such as roads and buildings is not consistently collected across the African continent. 

The purpose of this project is to overcome this data problem, by applying machine learning and artificial intelligence tool to a vast amount of unstructured data from daytime satellite imagery. Ultimately, this project aims to go beyond the use of nightlight luminosity as a proxy for economic development data and use high resolution, daytime satellite imagery to predict key infrastructure variables at national and subnational levels for less developed countries like in Africa. Daytime images contain more information about the landscape that is correlated with economic activity, but the images are highly complex and unstructured, making the extraction of meaningful information from them rather difficult. Our approach builds upon and further expands the work of \cite{Jean16}. The standard approach in the literature is to learn a representation out of  satellite images, that allow an interpretation of pixel activation that are important for predicting night time light or other target. This representation is then used to predict an aggregated wealth index. Instead, we directly predict infrastructure measures on the ground, albeit knowing that there is a wide spread scarcity of ground truth data.

\section{Motivation and Related Works}
 
Existing solutions for policy makers in developing countries often rely from traditional data gathering processes (i.e. surveys), which are costly and infrequent. Given the high costs, this data does not cover an entire country but only a sub-sample of geographic units. Our solution provides a low-cost method to collect valuable insights about economic development for every location in a country. Our methodology provides relevant decision makers in developing countries as well as NGOs and international organizations with very accurate counts of buildings and the length of roads for an entire country and continent.
For example, accurate building counts and density can be used in natural hazard preparedness tools as an indicator for an area’s vulnerability against natural disasters. Information about roads and settlement helps infrastructure agencies to quickly identify areas that lack market access, a key determinant for economic growth in developing countries.

Although relatively new, recent studies have begun to use different daytime satellite images to conduct novel economic research  \cite{donaldson2016}. Daytime images contain more information than night-time images and are thus a good alternative data source for empirical economics. \citet{Marx2013} used daytime images to analyse the effects of investment on housing quality in the slums of Kibera, Kenya. Investment was calculated based on the age of a household’s roof. The results showed that ethnicity plays an important role in determining investment in housing and belonging to the same tribe as that of the local chief has a positive effect on household investment. \cite{Engstrom17} used daytime satellite imagery and survey data to estimate the poverty rates of 3,500 km2 subnational areas in Sri Lanka.  Using a convolutional neural networks algorithm, they identify object features from raw images that were predictive of poverty estimates. The features examined by the study included built-up areas (buildings), cars, roof types, roads, railroads and different types of agriculture. The results showed that built-up areas, roads and roofing materials had strong effects on poverty rates. A suite of related work has used satellite images to predict population density \citep[e.g.][]{Simonyan15,Doupe16}, urban sprawl \cite{burchfield06}, urban markets \cite{baragwanath2019} electricity usage \cite{Robinson17}, as well as income levels \cite{Pandey2018}. More broadl, we also relate growing body of literature that uses other passively collected data to measure local economic activity \citep[e.g.][]{abelson2014,Blumenstock:2015,Chen2011,Henderson2012,Hodler2014}, Methodologically, our paper contributes to the large remote-sensing literature that applies high-dimensional techniques to extract features from satellite imagery \citep[e.g.][]{Jean:2016jm,Jean:2019ti,yeh2020using,ron15}.

\section{Data}
In general, reliable data at a more granular spatial level is very scarce for the African continent. This poses a particular challenge if the researcher wants to apply machine learning tools that require some form of ground truth data.

To overcome this problem, we accessed data from two open-data sources. The first one is Open Street Map, a collaborative project allowing volunteers around the world to contribute georeferenced information in an open-source GIS. We utilized \url{http://download.geofabrik.de/} to retrieve a complete snapshot of all geo-located objects Africa in 2018. In general, OSM coverage for Africa is very sparse and often non-existent outside urban areas. Our strategy to mitigate this issue, was to build an iterative procedure that would help us select areas (1 $\times$ km) with good  OSM coverage. We were then  able  to convert the geometric OSM data into an image mask.

\begin{figure}
\centering
  \includegraphics[width=.45\columnwidth]{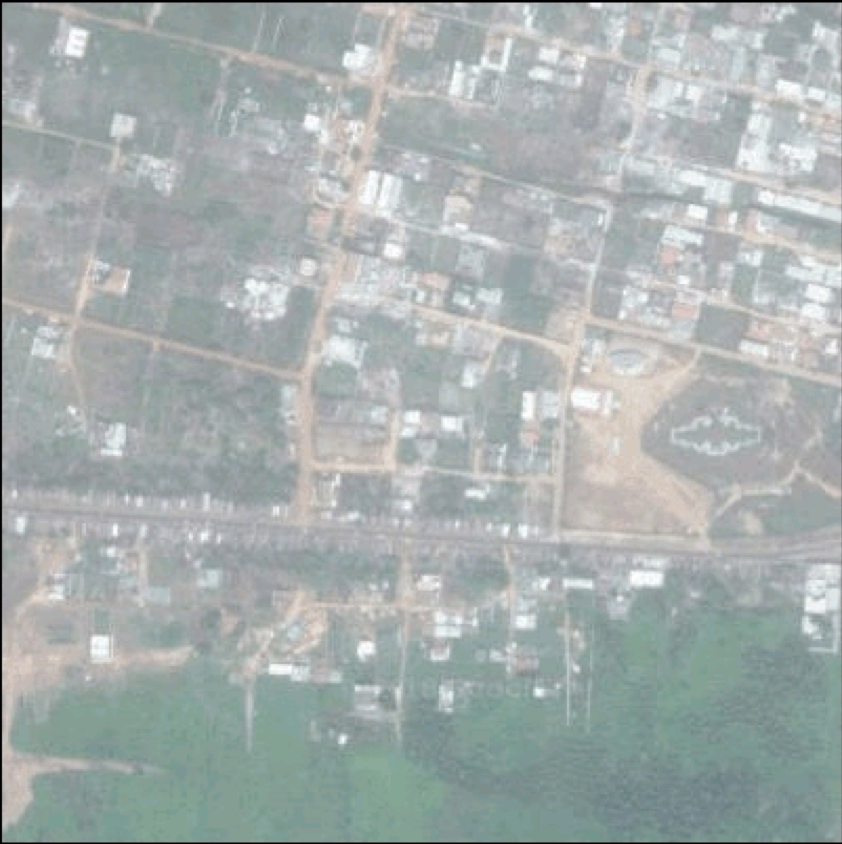}
  \includegraphics[width=.45\columnwidth]{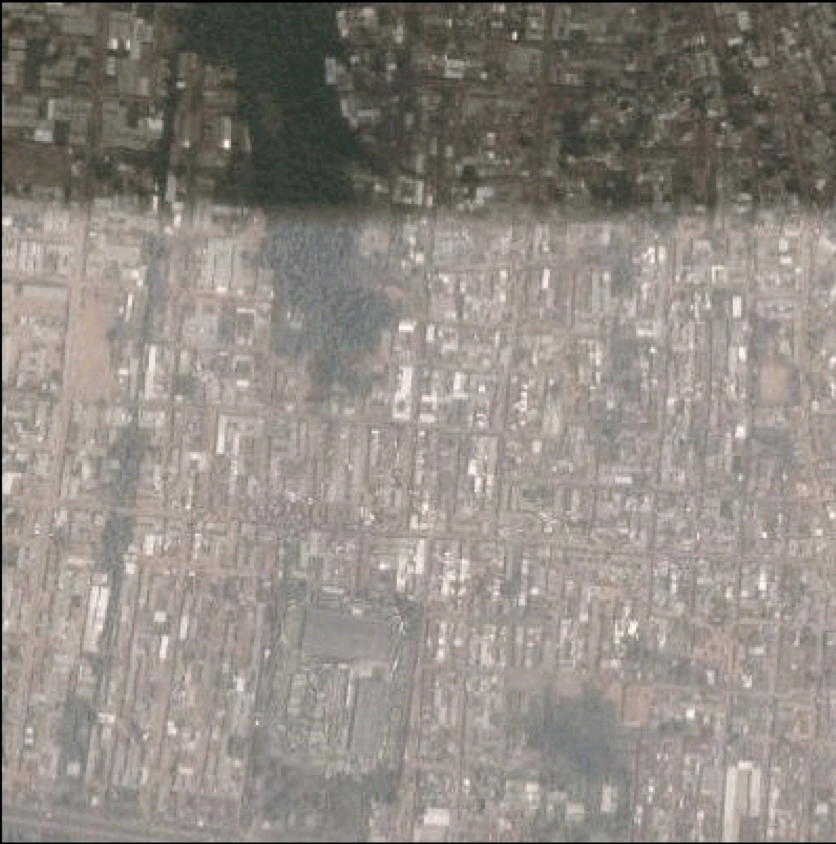}
  \caption{Quality: Freely available Africa imagery extracted via the Google MAPS api}
  \label{fig:africa_img}
\end{figure}

Our image data was collected in 2018 via the google maps api following the exact procedure as in \citet{Jean:2016jm}. This data set has be used in various studies \citep[e.g.][]{Jean:2019ti,Sheehan:2019iz,Uzkent:2019uk,Oshri:2018jf}. Again the same pattern as with OSM data emerges, the image quality of these freely available African images is not as good as in other places around the world, see figure \ref{fig:africa_img}. 

In the absence of reliable ground truth data, we selected the architecture based on data that we could make look like as if it would be from our target domain. For buildings, we employ imagery collected by drones in Africa from the "Open Cities AI Challenge: Segmenting Buildings for Disaster Resilience"\footnote{\url{https://www.drivendata.org/competitions/60/building-segmentation-disaster-resilience/}}, with the corresponding ground truth data provided and re-scaled and blurred the drone imagery. For roads, we build a model to select images with almost complete masks, albeit having missing roads and errors.  

We benchmark our proposed methodology against the latest publication of poverty predictions in Africa using their provided wealth index based on DHS cluster data \cite{yeh2020using}. As it is common in this literature an index is created with a principal component analysis (PCA) out of survey respondents. Again, due to data limitations, research in this area always only performed a in-sample validation. A true out of sample comparison would require a strict separation between the train and test set, something that is not possible if the PCA is calculated over all data points across all countries and therefore inflating the prediction results.  As such, \cite{yeh2020using} also provided an index that is based on within country survey respondents. This enables us to benchmark against both indices.

\begin{figure*}[h!]
  \centering
  \includegraphics[width=\linewidth]{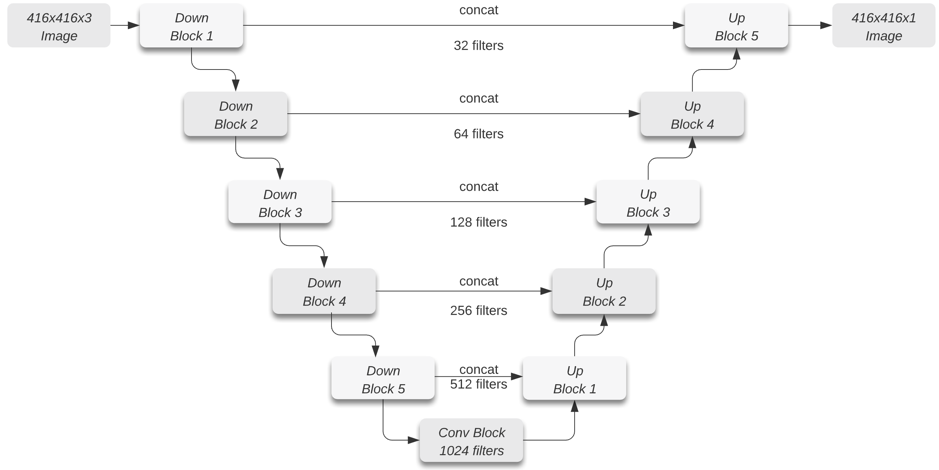}
  \caption{Satellite segmentation network (Sat-Unet) general overview. }
  \label{fig:model_over}
\end{figure*}

\section{Model Architecture}

\begin{figure}
  \centering
  \includegraphics[width=0.8\linewidth]{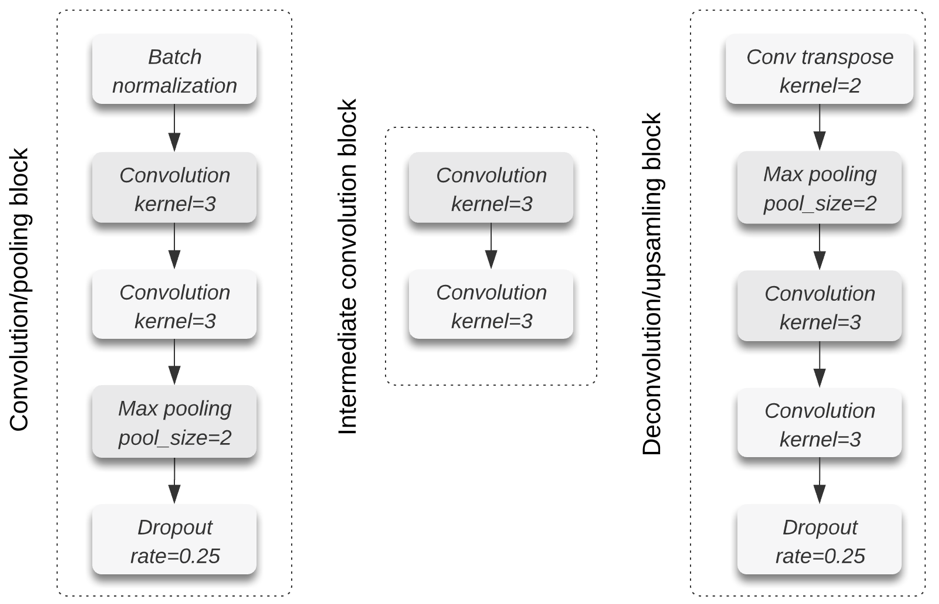}
  \caption{Blocks of the segmentation network}
  \label{fig:blocks}
\end{figure}

In principle, we follow the outline of the well known U-Net architecture for medial images \cite{ron15} and modify it for satellite images creating a Satellite-U-Net (Sat-Unet). figure \ref{fig:model_over} provides a general overview of our approach. The network contains 61 layers in total, with 11 major blocks of 3 types: convolution / down-sampling block, intermediate convolution block and the de-convolution/up-sampling block. The convolution block, shown in figure \ref{fig:blocks}, consists of a batch normalization layer, two convolution layers with the kernel of (3,3) and a dropout layer. The dropout layer is not used in the first down-sampling block. The number of filters in down-sampling blocks (encoder part) starts from 32 and doubles every time in the following block reaching 1024 in the intermediate convolution block, and then decreases in the up-sampling blocks (decoder) with the coefficient 0.5. The core difference from \cite{ron15} is that instead of up-sampling layers, we are using transposed convolution layers,  which performs the reverse convolution operation \cite{conv}. In addition, we added drop-out layers after each convolution and de-convolution block.

In a 400x400 image the number of pixel belong to a house or a road, class one, vs the number of pixels of zero-class (non classified space) is up to $10^4$ times higher, creating a severe class imbalance. We address this issue with a hybrid loss function. First, we use the loss of the sum of binary cross entropy

\begin{equation}
      H_p(q) = -\frac{1}{N} \sum^N_{i=1} y_i \cdot log(p(y_i))+(1-y_i)\cdot log(1-p(y_i)) 
\end{equation}
and the Sorensen-Dice coefficient:
\begin{equation}
  S_V = \frac{2|a\cdot b|}{|a|^2+|b|^2}\
\end{equation}
combined. As metric we used the intersection over a union (the Jaccard index)
\begin{equation}
  J(A,B) = \frac{|A\cap B|}{|A\cup B|} = \frac{|A\cap B|}{|A|+|B|-|A\cap B|}\
\end{equation}

\subsection{Data pre-processing}
To make image input size is 400x400, compatible with a factor of 32  to conform the shape reduction coefficients of the network, we added a padding of 8. On average across our images, the RGB colours maximum was around 180-190 out of the maximum of 255. Color channels re-scaling has been implemented to intensify colors before feeding the image into the network. For augmentation we used rotation by 90, 180 and 270 degrees.

\subsection{Model evaluation under uncertainty}

The main difficulty in choosing the exact architecture for the road network was the lack of a sufficiently large amount of, error-free, ground truth data. Therefore, we used the following iterative strategy:
\begin{enumerate}
\item Create an initial mask with OSM data and train on them.
\item Filter out masks, where the model predicts significantly more objects than the OSM mask has.
\item Retrain the model on the filtered data-set.
\end{enumerate}
Due to the large possible set of images to train from, around 22 million, we first selected a subset based on OSM data. As our main focus is to get an indicator of the economic development, the best case would be to find areas of economic activity. OSM has a classification for commercial buildings, which is rarely used (2143/22 mil). We selected areas in the same ADM2 regions of those images based on descending order of square meters occupied by buildings on a uniform grid, until we had selected a base set of 10000 masks. Next, we trained our Sat-Unet model for roads on all masks as labels we had created from OSM. 

\begin{figure}[h!]
  \centering
  \includegraphics[width=\linewidth]{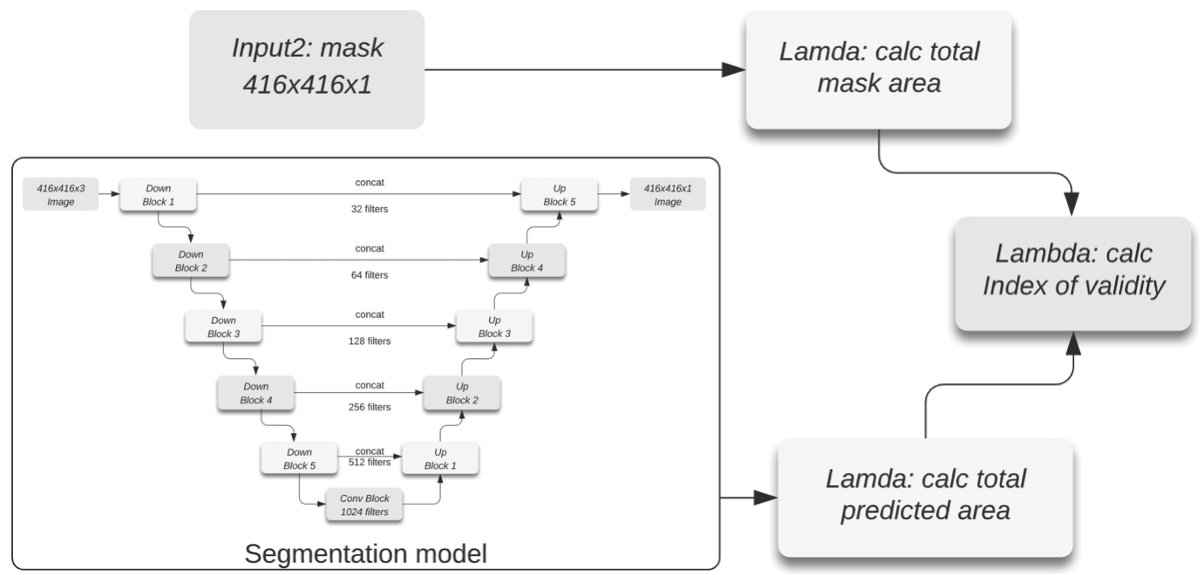}
  \caption{The model judge}
  \label{fig:judge}
\end{figure}

\begin{figure*}[h!]
\centering
  \includegraphics[width=.31\textwidth]{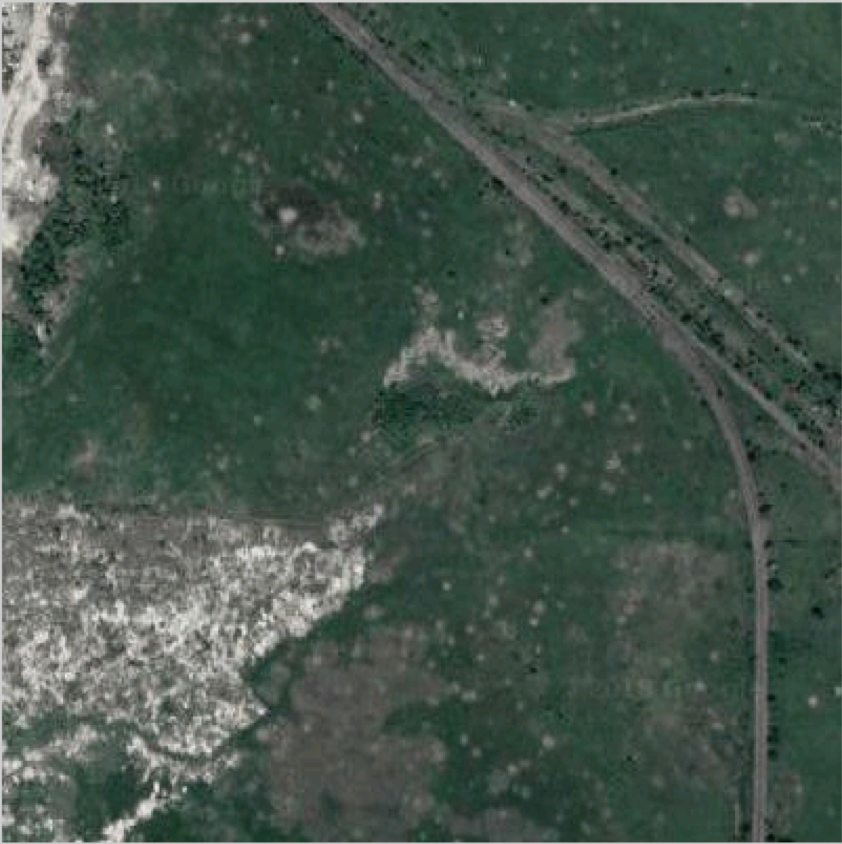}
  \includegraphics[width=.31\textwidth]{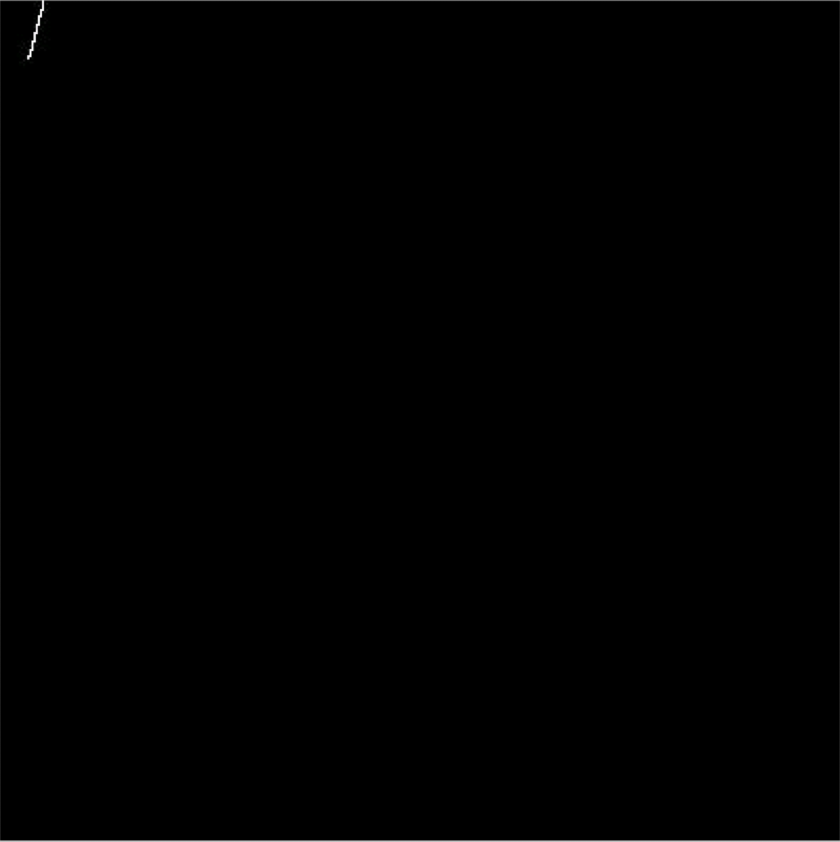}
  \includegraphics[width=.31\textwidth]{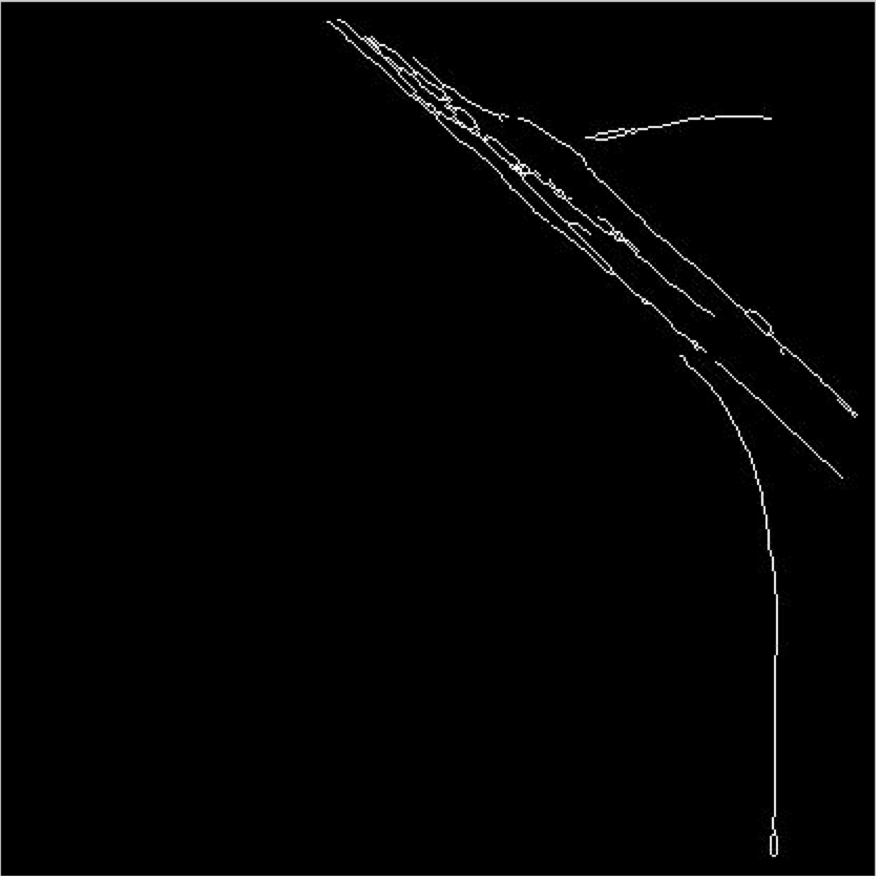}
  \caption{The filter removes incomplete OSM masks. From left to right: original image, OSM mask, prediction of the trained model on the stage 1.}
  \label{fig:filt}
\end{figure*}

\begin{figure}
\begin{center}
       \includegraphics[width=0.58\columnwidth]{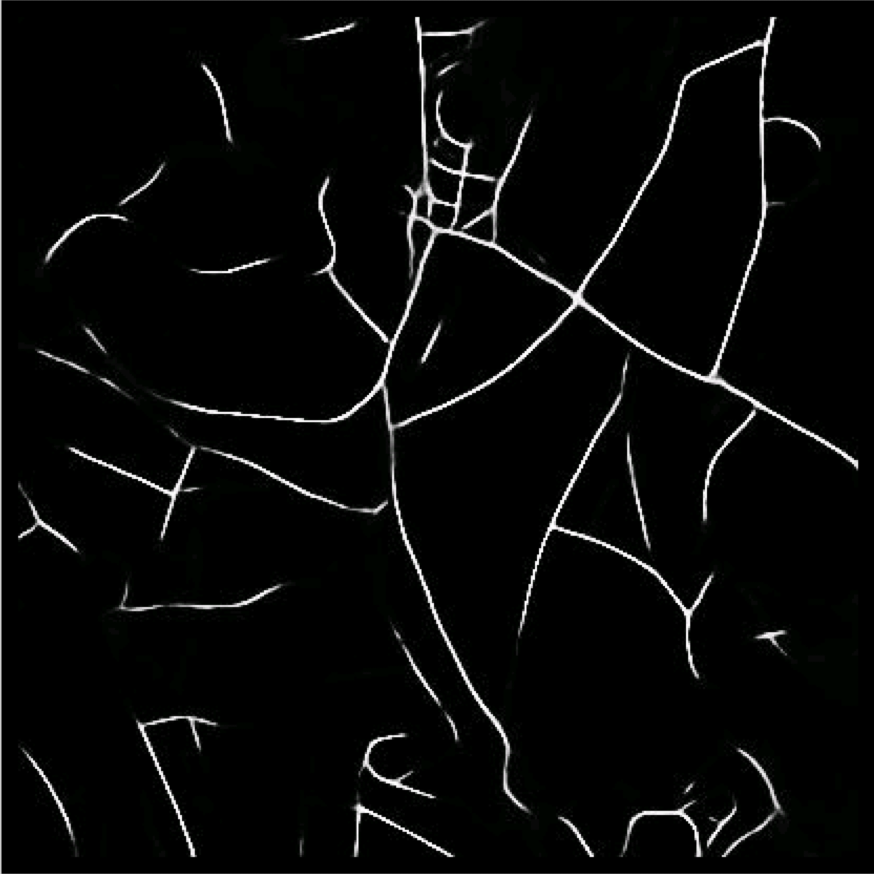}%
    \end{center}
    \begin{center}
      \includegraphics[width=0.58\columnwidth]{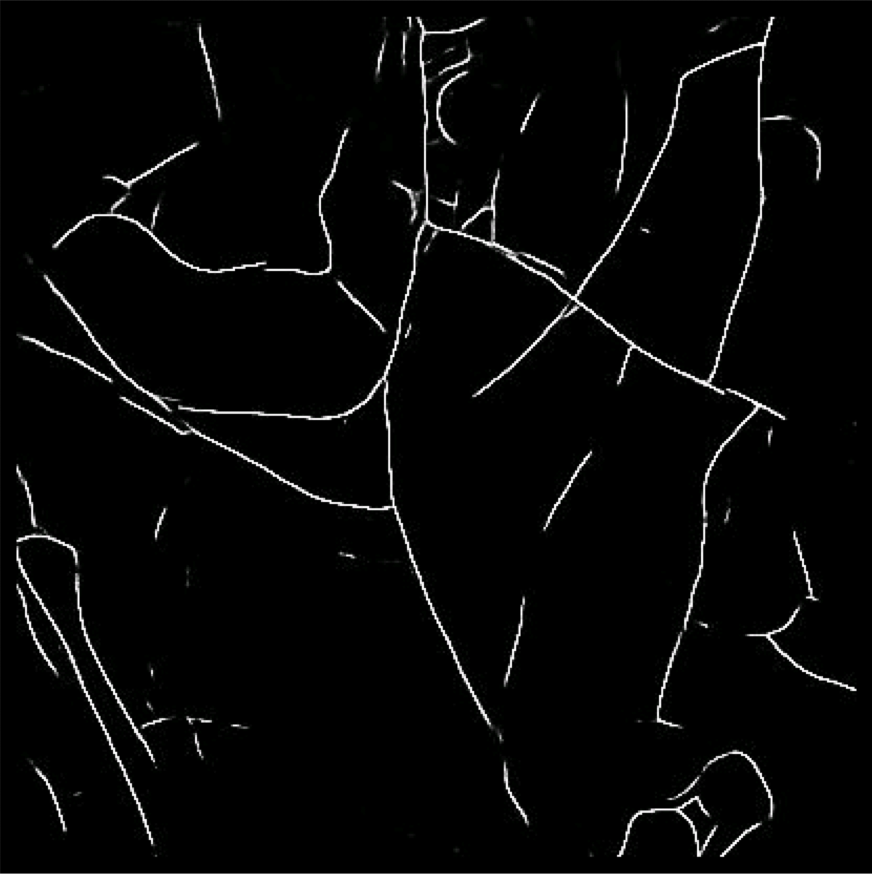}
    \end{center}
    \begin{center}
\includegraphics[width=0.58\columnwidth]{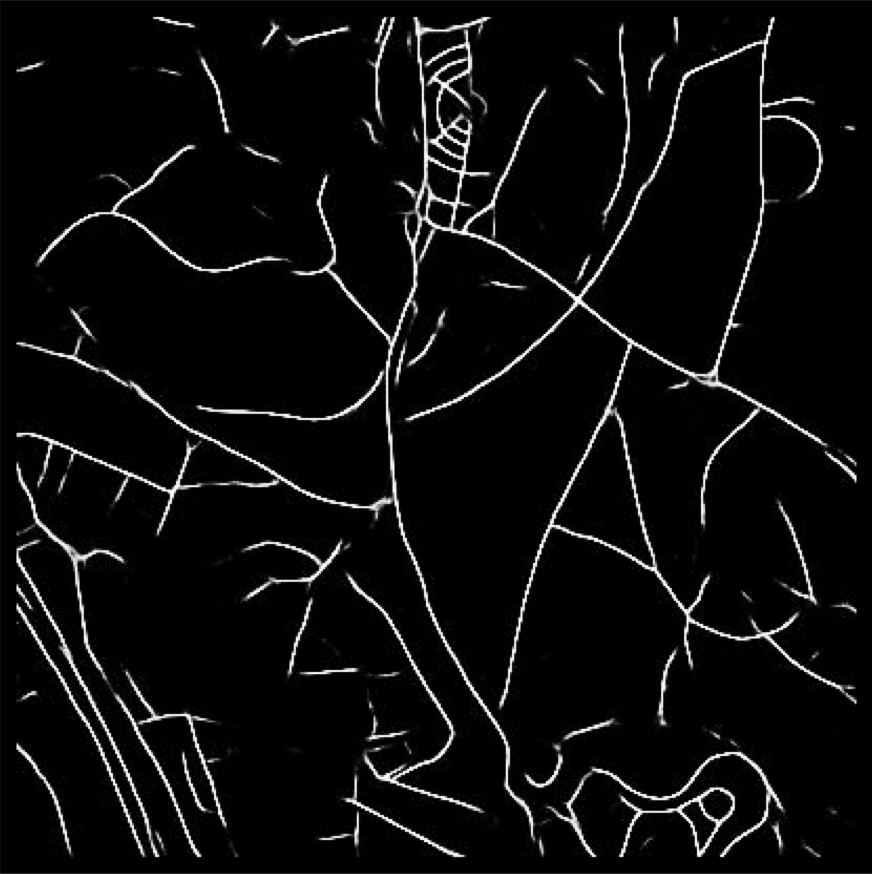}%
    \end{center}
    \begin{center}
\includegraphics[width=0.58\columnwidth]{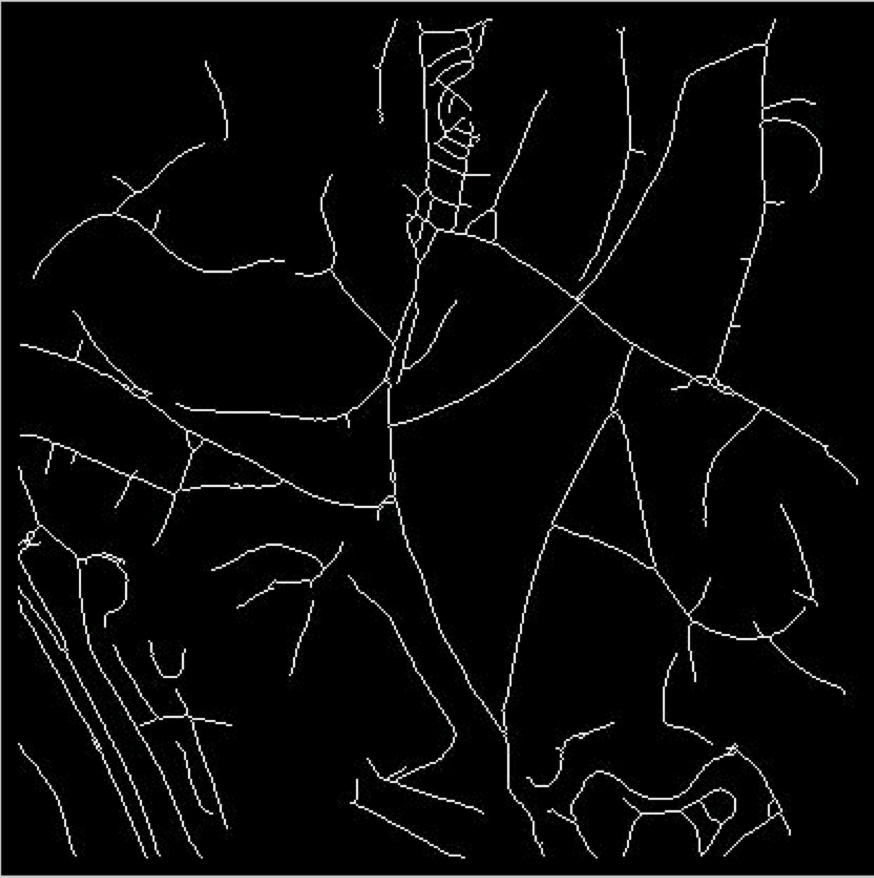}
    \end{center}
  \caption{Ensembling: The first three are predicted masks based on different random seeds, the last one is the resulting mask.}
  \label{fig:ensem}
\end{figure}

\paragraph{The Judge:} For filtering purposes the Sat-U-Net based model Judge has been created with an additional input for the OSM mask. Using transfer learning the weights of the pretrained Sat-UNet model have been transferred to the bottom layers of the Judge for the mask creation from the original image, and top layers perform the calculations of the Index of validity, using the calculated mask and the OSM mask as inputs. Combining everything in a single GPU model allows to achieve more than 50x increase in performance comparing to CPU-based technique. \ref{fig:judge}. The index of validity is
\begin{equation}
  \alpha = \frac{\sum^K_{i=1} i}{\sum^K_{j=1} j}\
\end{equation}
where i,j- pixel values of predicted mask and the OSM respectively. The resulting filtering model decreased the data-set approximately by 40\%, filtering out instances like those presenting in figure \ref{fig:filt}.
Furthermore, as our network predictions had very low values caused by model uncertainty, we re-trained several Sat-Unet model on the selected images with different random seeds. This ensemble learning significantly increased our predictive performance, as shown in figure \ref{fig:ensem}. The top three are predictions on the test set, while the last image is the combination of all three with the pixels reduced to the skeleton for counting.

\section{Building Recognition}
For building recognition,  we used Open Cities AI Challenge data set as ground truth data set. This data set contains imagery of several African cities in a ultra high resolution of up to 5cm per pixel. Each city is split by square areas and for each image there is a GeoJSON file with vector data describing contours of the buildings. Out of these geometric data, we created a contour layer and a centroid layer, which represents the center of every building structure. We down scaled the images to the scale of 1.53 pixel/meter to match Google Maps images and split the images into 400$\times$400 patches. As target mask for training we used the centroid layer as we were interested in the number of houses, not necessary the full area extent.

\begin{figure}[h!]
\centering
  \includegraphics[width=.45\columnwidth]{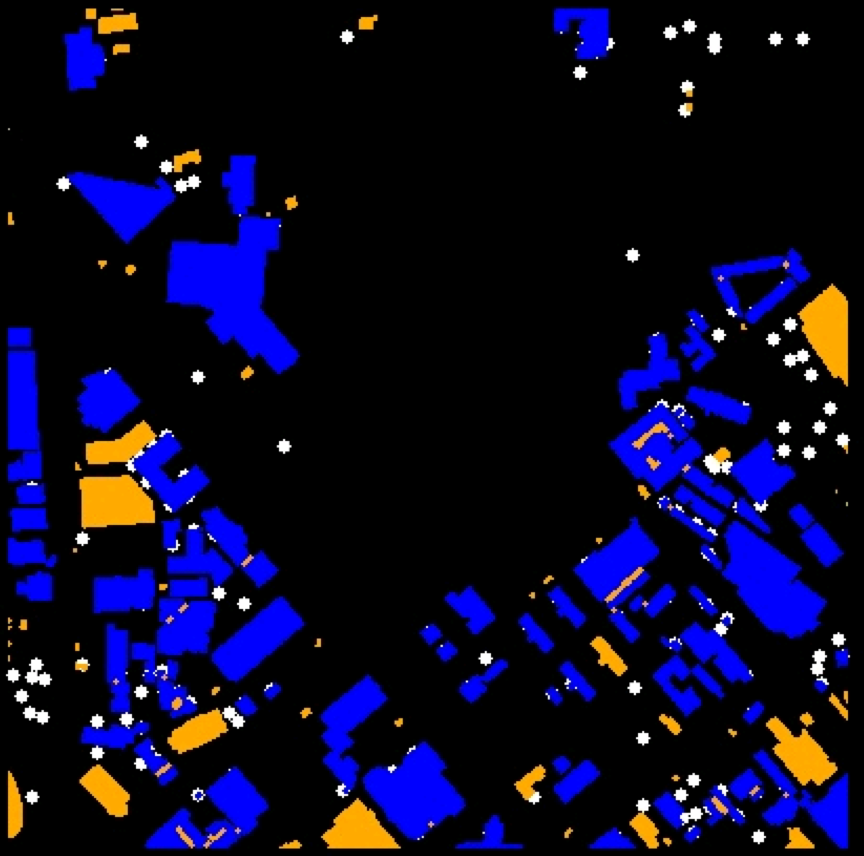}
  \includegraphics[width=.45\columnwidth]{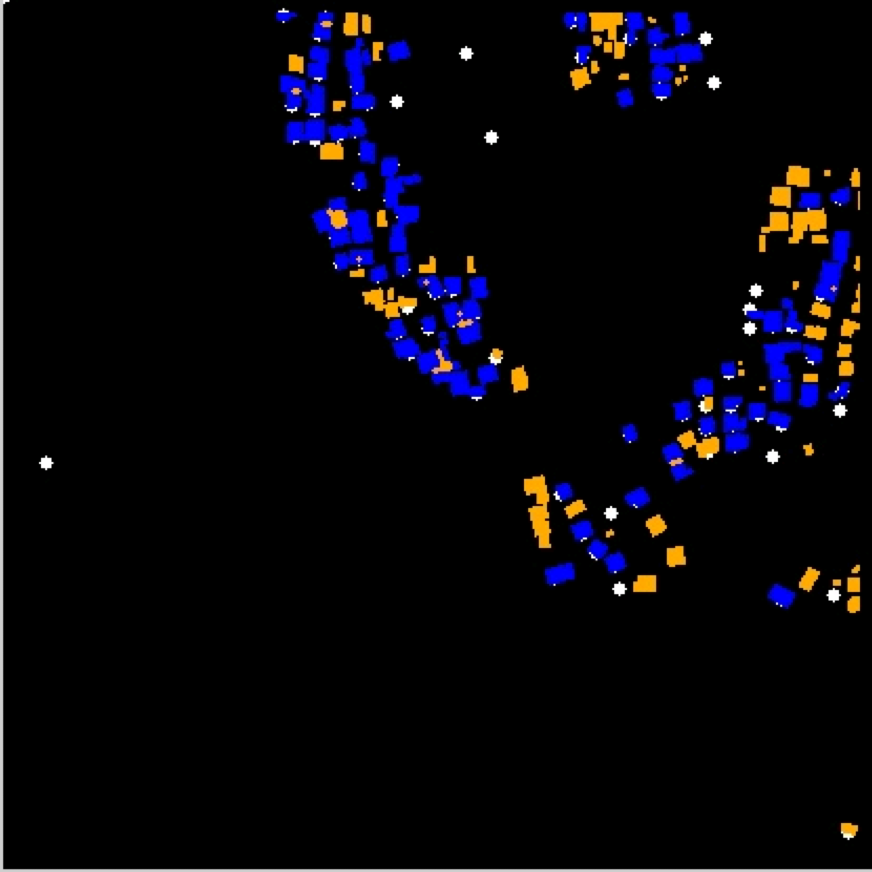}
  \caption{Africa, building model stage, contour-in-contour evaluation. Buildings that were predicted correctly are in blue, not predicted in orange. Orange areas of random shapes inside building blocks are usually courtyard areas and not considered as a wrong prediction.}
  \label{contour}
\end{figure}

This high-quality ground-truth data further allowed us to experiment with different architectures. We replaced the encoder part in the Sat-Unet model \ref{fig:model_over} with the inception v3\cite{xia2017inception} and the resnet50\cite{res15}. In every instance we reset all the weights to random before training but we did not make use of any transfer learning.  Table \ref{tab:modpresel} presents the evaluation results on our test set. We also compared how well the network performs in counting the correct house based on the jaccard index, visually shown in figure \ref{contour}.

\begin{table}
\centering
\begin{tabular}{lrrr}
\hline
Type     & Loss  & IOU   & Dice  \\ \hline
Sat-Unet     & 1.299 & 0.010 & 0.019 \\
Incep3-Unet   & 0.895 & 0.282 & 0.329 \\
Resnet50-Unet & 0.870 & 0.326 & 0.380 \\ \hline
\end{tabular}
\caption{ Model pre-selection results, an encoder of ResNet50 has been chosen.}
\label{tab:modpresel}
\end{table}

The performance of the Incep3-Unet and Resnet50-Unet was quite similar. For the final model selection we trained both architectures on the previous selected masks out of OSM, and compared their performance in terms of their predictability on a test set. Table \ref{tab:gsres} demonstrates the effect of different thresholds on the performance.The threshold is in color intensity units (range 0-255), TP is true positive, when the predicted centroid is located inside the building contour of the mask, Pred-To-Mask coefficient is the ratio between predicted number of houses and the ground truth number, and False Positives (FP): 
\begin{equation}
     FP \gets 100 \cdot \frac{TotalPred-TP}{TotalPred}
\end{equation}
A threshold of 15 has the closest Prediction-to-mask score, acceptable TP and FP rates. Therefore, we picked this threshold for further modeling. The final model Resnet50-Unet has 268 layers. To avoid the vanishing gradient problem with the depth of hundreds of layers ResNet uses skip connections, it adds input information of the convolution block to its output. In addition skip connections give the model the ability to learn the identity function which guarantee the similar performance of the lower and higher layers \cite{res15}.

\begin{table}
\centering
\begin{tabular}{lrrrr}
\hline
Type  & Thresh & TP      & Pred-To-Mask & FP     \\ \hline
Incep3-Unet  & 05     & 47.876  & 126.226      & 61.143 \\
Incep3-Unet  & 10     & 45.639  & 112.446      & 60.903 \\
Incep3-Unet  & 15     & 44.243  & 102.651      & 61.529 \\
Incep3-Unet  & 25     & 42.162  & 85.075       & 61.360 \\
Resnet50-Unet & 05     & 50.640  & 129.173      & 63.654 \\
Resnet50-Unet & 10     & 48.496  & 113.195      & 61.774 \\
Resnet50-Unet & 15     & 47.1438 & 105.009      & 61.644 \\
Resnet50-Unet & 25     & 45.099  & 93.966       & 61.750 \\ \hline
\end{tabular}

\caption{\label{tab:gsres} Threshold effect}
\end{table}

\section{Results}

\subsection{Benchmark Results}
We compare our prediction results of buildings and roads to the latest benchmark study in the field of poverty prediction based on DHS data\cite{yeh2020using}. The DHS data is collected in various waves across countries and years. For the comparison use of the most recent wave available for a country and use the aggregated wealth index data. Two indices are provided, the first \emph{wealthpooled} is the PCA calculation across all years, while the second index \emph{wealth}, is calculated out of the survey respondents only within a country and year, allowing out of sample comparison.

\cite{yeh2020using} did not provide any out of sample estimates, therefore we fully replicated their method by getting all their images they used for their location and trained  their combined CNN model of multi-spectrum and night lights images to determine economic well-being in Africa, with \emph{wealth} as label for the last of their training folds (D). The performance results are almost identical in terms of R-squared. In their study they also found a high correlation of the measures to other type's of aggregation, such the sum total of all assets. Using their \emph{wealthpool} predictions as predictor for \emph{wealth}, the r-squared is around 0.61 vs 0.67 for \emph{wealthpool}.

\subsection{Prediction Results}
The DHS location data from \cite{yeh2020using} has 7,315 unique cluster location in the last wave of each respective country. We use a 5km radius corresponding to the possible displacement of survey measurements, as selection criteria  to select our images. For every square km we predict the number of buildings, the number of roads as well as calculate the night time light\cite{elvidge2017viirs} by grid cell. We then aggregate the roughly 1.5 million images into features, by building the sum, averag and quantiles by cluster across all 3 input variables. In total, this leaves us with 6,112 locations. We perform LOOCV cross validation by country, as we are interested in predicting the marginal unit if we would use the model to get data for one extra country. We also iterate over standard machine learning algorithms without any hyper parameter tuning, by only using the default settings. Table \ref{Wealth_benchmark} and table \ref{Wealthpooled_benchmark} present the results for out of sample and out of country predictions, respectively. As expected, using a normalized outcome measures across all samples, inflates the performance. In comparison to the previous literature, our predictions  show an increased predictive performance, both in and out of sample.

\begin{table*}
\centering
\begin{tabular}{lccccc}
\hline
Model       & Buildings & Roads & Buildings \& Roads & Nighttime light & Buildings, Road \& \ Nighttime light \\ \hline
ridge       & 0.322     & 0.154 & 0.155              & 0.353           & 0.155                                \\
r-tree      & 0.336     & 0.459 & 0.468              & 0.540           & 0.567                                \\
r-tree (bs) & 0.437     & 0.552 & 0.576              & 0.637           & 0.652                                \\
r-tree (bg) & 0.495     & 0.609 & 0.639              & 0.678           & 0.723                                \\ \hline
\end{tabular}

\caption{Predictive performance of satellite predictions, r-squared based on LOOCV on out of country and out of sample predictions by country for \emph{wealth}}
\label{Wealth_benchmark}
\end{table*}

\begin{table*}
\centering
\begin{tabular}{lccccc}
\hline
Model       & Buildings & Roads & Buildings \& Roads & Nighttime light & Buildings, Road \& Nighttime   light \\ \hline
ridge       & 0.110     & 0.146 & 0.146              & 0.311           & 0.147                                \\
r-tree      & 0.239     & 0.401 & 0.402              & 0.589           & 0.594                                \\
r-tree (bs) & 0.359     & 0.494 & 0.525              & 0.678           & 0.692                                \\
r-tree (bg) & 0.419     & 0.563 & 0.595              & 0.719           & 0.738                                \\ \hline
\end{tabular}

\caption{Predictive performance of satellite predictions, r-squared based on LOOCV on out of country predictions by country using \emph{wealthpooled}}
\label{Wealthpooled_benchmark}
\end{table*}

\section{Discussion and Future Work}

This paper introduces a novel and scalable method to predict road and housing infrastructure from daytime satellite imagery. Compared to existing approaches, we achieve higher predictive performance by training a U-net style architecture using ground-truth data from a subset of images. Using satellite images from 21 African countries we show how our method can be used to generate very granular information about the stock of housing and road infrastructure for regions in the world, where reliable information about the local level of economic development is hardly available. Consistently measured and comparable indicators about local economic development are crucial inputs for governments in developing countries as well as international organizations in their decision where to allocate scarce public funds and development aid. 

The predictions generated by our method can be directly included in existing decision support systems. For example, international organization such as the Red Cross are using similar data at the local level to evaluate an area's vulnerability against natural hazards. Our data can be considered as more granular complements to existing measures of the local stock of physical infrastructure. Numerous charitable organizations already rely on satellite imagery to identify districts of African countries that are among the least developed \cite[e.g.][]{abelson2014}. Our approach provides a low-cost and scalable alternative to identify areas that are in need. In addition, the Open Street Map mapping community would benefit from our findings as well. The road prediction model could be used worldwide to help completing the road network or help narrowing down possible errors in the data.

Finally, our approach is an important methodological contribution to the large group of scholars from varying disciplines working in the area of poverty measurement. The majority of the existing research focuses on predicting poverty based on aggregate household wealth. This paper shows that predicting poverty measures can also be viewed as a simple high dimensional feature representation problem. Our study is a proof-of-concept exercise to show that combining daytime satellite imagery, open source ground truth data and machine learning tools can translate unstructured image data into valuable insights about local economic development at an unprecedented scale.

\section{Appendix}
List of countries used: Angola, Benin, Burkina Faso, Cameroon, Central African Republic, Côte d'Ivoire, Democratic Republic of the Congo, Ethiopia, Ghana, Guinea, Kenya, Lesotho, Malawi, Mali, Nigeria, Rwanda, Senegal, Sierra Leone, Tanzania, Togo and Uganda

\bibliography{main_aaai}

\end{document}